\documentclass[12pt]{article}
\usepackage{graphicx}
\usepackage{color}
\usepackage{natbib}
\usepackage{bm,booktabs,physics}
\makeatletter
\def\maxwidth{ %
  \ifdim\Gin@nat@width>\linewidth
    \linewidth
  \else
    \Gin@nat@width
  \fi
}
\makeatother
\usepackage{setspace}
\usepackage{caption}
\usepackage[margin=1in,lines=25]{geometry}
\definecolor{fgcolor}{rgb}{0, 0, 0}
\newcommand{\hlnum}[1]{\textcolor[rgb]{0.533,0,0.133}{#1}}%
\newcommand{\hlstr}[1]{\textcolor[rgb]{0.667,0.267,0}{#1}}%
\newcommand{\hlopt}[1]{\textcolor[rgb]{0,0,0}{\textbf{#1}}}%
\newcommand{\hlstd}[1]{\textcolor[rgb]{0,0,0}{#1}}%
\newcommand{\hlkwa}[1]{\textcolor[rgb]{0.4,0.067,0.067}{\textbf{#1}}}%
\newcommand{\hlkwb}[1]{\textcolor[rgb]{0,0,0.4}{\textbf{#1}}}%
\newcommand{\hlkwc}[1]{\textcolor[rgb]{0,0,0.4}{#1}}%
\newcommand{\hlkwd}[1]{\textcolor[rgb]{0,0.267,0.4}{#1}}%

\usepackage{framed}
\makeatletter
\newenvironment{kframe}{%
 \def\at@end@of@kframe{}%
 \ifinner\ifhmode%
  \def\at@end@of@kframe{\end{minipage}}%
  \begin{minipage}{\columnwidth}%
 \fi\fi%
 \def\FrameCommand##1{\hskip\@totalleftmargin \hskip-\fboxsep
 \colorbox{shadecolor}{##1}\hskip-\fboxsep
     \hskip-\linewidth \hskip-\@totalleftmargin \hskip\columnwidth}%
 \MakeFramed {\advance\hsize-\width
   \@totalleftmargin\z@ \linewidth\hsize
   \@setminipage}}%
 {\par\unskip\endMakeFramed%
 \at@end@of@kframe}
\makeatother

\definecolor{shadecolor}{rgb}{.97, .97, .97}
\definecolor{messagecolor}{rgb}{0, 0, 0}
\definecolor{warningcolor}{rgb}{1, 0, 1}
\definecolor{errorcolor}{rgb}{1, 0, 0}
\newenvironment{knitrout}{}{} 
\newcommand{\ntt}{\normalfont\ttfamily}

\newcommand{\pkg}[1]{{\protect\ntt#1}}

\usepackage{alltt}
\newcommand{\blue}[1]{{\textcolor{blue}{#1}}}

\newcommand{\ttt}[1]{\blue{\texttt{#1}}}
\usepackage[colorlinks, citecolor=blue, linkcolor=blue]{hyperref}
\usepackage{amsmath,amsthm,amssymb}
\begin{document}

\title{A comparison of Bayesian accelerated failure time models with spatially
varying coefficients
}


\author{Guanyu Hu$^*$~~~~Yishu Xue\thanks{Department of Statistics, University of Connecticut, Storrs, CT 06269}~\thanks{Correspondence: \url{yishu.xue@uconn.edu}}~~~~Fred Huffer\thanks{Department of Statistics, Florida State University, Tallahassee, FL 32306}}

\date{\today}

\maketitle

\begin{abstract}
The accelerated failure time (AFT) model is a commonly used tool in analyzing
survival data. In public health studies, data is often collected from medical
service providers in different locations. Survival rates from different
locations often present geographically varying patterns. In this paper, we
focus
on the accelerated failure time model with spatially varying coefficients. We
compare a three different types of the priors for spatially varying
coefficients. A model selection criterion, logarithm of the pseudo-marginal
likelihood (LPML), is developed to assess the fit of AFT model with different
priors.  Extensive simulation studies are carried out to examine the empirical
performance of the proposed methods.  Finally, we apply our model to SEER data
on prostate cancer in Louisiana and demonstrate the existence of spatially
varying effects on survival rates from prostate cancer. \\

\end{abstract}

\section{Introduction}\label{ch3Introduction}

Patient data in public health studies is often collected on certain
administrative divisions such as counties or provinces. Oftentimes, the patient
group in different regions have similar characteristics yet exhibit different
patterns of survival outcomes, which leads us to investigate in the
geographical
variation of covariate effects.

There is much recent work analyzing geographical patterns of survival data. For
example, \cite{henderson2012modeling} used the proportional hazards model to
model
spatial variation in survival of leukemia patients in northwest England; 
\cite{banerjee2005semiparametric} applied a spatial frailty model to
infant mortality in Minnesota by using geostatistical or Gaussian Markov random
field priors for the spatial component; \cite{zhou2008joint} applied the
conditional autoregressive (CAR) model in a parametric survival model to
construct a joint spatial survival model for prostate cancer data, and
\cite{zhang2011bayesian} modeled the spatial random effects in an
accelerated failure rate (AFT) model using the CAR prior. In all these works,
spatial variation is modeled as spatial random effect, while the variation
in the covariate effects for risk factors are not accounted for.
From the spatially varying coefficients perspective, 
\cite{gelfand2003spatial} proposed a model that allows the coefficients in a
regression model to vary at the local or
subregional level by viewing them as realizations of a Gaussian process
with a certain covariance structure that is decided by the relationship
between spatial locations.
 \cite{reich2010bayesian,boehm2015spatial} applied spatially
varying coefficients in a generalized linear model to investigate the health
effects of fine particulate matter components. An application of the spatially
varying coefficients methodology to the Cox model was proposed in
\cite{xue2019geographically} from the frequentist perspective.

In this work, we propose a Bayesian AFT model with spatially varying
coefficients. Specifically, the variation in coefficient vectors is modeled
using three different priors: Gaussian, CAR, and the Dirichlet process (DP),
corresponding to different possible true underlying variation patterns.
A model selection criterion, logarithm of the pseudo-marginal likelihood
(LPML), is employed to assess the fitness of three different priors. Our
simulation
studies showed promising empirical performance of the proposed methods in
both non-spatially varying and spatially varying cases. Furthermore, our
proposed criterion also select the best fitness model. In addition, our
proposed Bayesian approach reveals interesting features of the prostate cancer
data for Louisiana.

The rest of this paper is organized as follows. In Section~\ref{sec:baft} we
propose the Bayesian AFT model with spatially varying coefficients for both
uni-covariate and multi-covariate cases. Three prior distributions used to
account for the spatial variation are introduced. In Section~\ref{sec:binf}, we
present the computation for the proposed model using the powerful \textsf{R}
package \pkg{nimble} \citep{de2017programming}, and propose the corresponding
model selection technique using the LPML. Simulation studies are conducted
in Section~\ref{sec:simu}, and we
illustrate its practical use in Section~\ref{sec:apply} using the prostate
cancer data for Louisiana from the SEER program. We conclude the paper with a
brief discussion in Section~\ref{sec:disc}.

\section{The Bayesian AFT}\label{sec:baft}
\subsection{AFT with Spatially Varying Coefficients}

Let $T_{\ell}(s_i)$ denote the survival time for patient $\ell$ at location
$s_i$, and $X_\ell(s_i)$ denote a covariate corresponding to $T_{\ell}(s_i)$,
where $i=1,2,...,n$, and $\ell=1,2,...,n_i$, with $n_i$ denoting the number of
the
patients at $s_i$. We propose the following spatial AFT model:
\begin{eqnarray}
  \log(T_{\ell}(s_i))=\tilde{\beta}_{0}(s_i)+X_\ell(s_i)\tilde{\beta}_1(s_i)+
  \sigma(s_i)\epsilon_{\ell}(s_i),
	\label{eq:singlemodel}
\end{eqnarray}
where $\tilde{\beta}_1(s_i)=\beta_1+\beta_1(s_i)$ is the slope at location
$s_i$, $\tilde{\beta}_0(s_i)=\beta_0+\beta_0(s_i)$ is the intercept at location
$s_i$, $\sigma(s_i)$ is the scale parameter at location $s_i$, and the
$\epsilon_{l}(s_i)$'s are i.i.d.\ random errors. The term $\beta_1(s_i)$ can be
regarded as a random spatial adjustment at location $s_i$ to the overall slope
$\beta_1$.

Let $f(t_\ell(s_i))$ be the density function of $T_\ell(s_i)$ and $f_0(\cdot)$
denote the density function of $\epsilon_\ell(s_i)$. Further, denote the
survival function of $T_\ell(s_i)$ as $S(t_\ell(s_i))$, and of
$\epsilon_\ell(s_i)$ as $S_0(\cdot)$. We consider right-censored survival
observations $(t_\ell(s_i),\delta_\ell(s_i))$, where
$\delta_\ell(s_i)=1(t_\ell(s_i)\leq C_\ell(s_i))$ with 1() being the indicator function,
$C_\ell(s_i)$ being the censoring time. Then the likelihood function can be
written as:
\begin{eqnarray}
  L=\prod_{i=1}^n \prod_{\ell=1}^{n_i} f(t_l(s_i))^{\delta_\ell(s_i)}
  S(t_\ell(s_i))^{1-\delta_{\ell}(s_i)},
	\label{single_likelihood}
\end{eqnarray}
where
\[
\begin{split}
  f(t_\ell(s_i))&=\frac{1}{\sigma(s_i)t_\ell(s_i)}f_0(\log(t_{i}(s_i))-
  (\tilde{\beta}_{0}(s_i)+X(s_i)\tilde{\beta}_1(s_i))/\sigma(s_i)),\\
  S(t_\ell(s_i))&=S_0(\log(t_{\ell}(s_i))-(\tilde{\beta}_{0}(s_i)+X(s_i)
  \tilde{\beta}_1(s_i))/\sigma(s_i))\,.  
  \end{split}
\]
In a parametric model, $S_0(.)$ can be assumed to be the standard normal
distribution, the standard extreme value distribution, or the logistic
distribution, etc.  They lead to different distributions (e.g., exponential
distribution, Weibull distribution, log-logistic distribution and log- normal
distribution) for the survival times $T_\ell(s_i)$ to complete the model
specification.

Now, instead of considering a single covariate, we consider a
$p$-dimensional covariate vector for each observation. Let
$\bm{X}_\ell(s_i)$ be the covariate vector for patient $\ell$ at location $s_i$, where
$\bm{X}_\ell(s_i)$ includes an initial $1$ (for the intercept). Equation
\eqref{eq:singlemodel} can be written into the following form:
\begin{equation}
\label{eq:multimodel}
  \log(T_{\ell}(s_i))=\bm{X}_\ell(s_i)\bm{\tilde{\beta}}(s_i)+
  \sigma(s_i)\epsilon_{\ell}(s_i),~~\ell=1,\ldots, n_i,~ i=1,\ldots, n.
\end{equation}

\subsection{Gaussian Process Prior}
The most straightforward prior for spatially varying coefficients is the
Gaussian process prior \citep{gelfand2003spatial}. The Gaussian process prior assumes that:
\begin{equation}
{\tilde{\bm{\beta}}}\mid \bm{\mu}_{\bm{\beta}},\bm{T} \sim \mbox{MVN}
(1_{n\times 1}\otimes
\bm{\mu}_{\bm{\beta}},\bm{H}(\phi)\otimes \bm{T})
\label{eq:gaussian}
\end{equation}
where ${\tilde{\bm{\beta}}}=(\tilde{\bm{\beta}}(\bm{s}_1)^\top,\ldots,\tilde{\bm
{\beta}}(\bm{s}_n)^\top)^\top$, $\bm{\mu}_{\bm{\beta}}$ is a $p \times 1$
vector,  $\bm{H}(\phi)$ is a $n\times n$ matrix of spatial correlations between
the $n$ observed locations, $\bm{T}$ is a $p \times p$ covariance matrix
associated with an observation vector at any spatial location, and $\otimes$
denotes the Kronecker product. The $(i,j)$th entry of $\bm{H}(\phi)$ is
$\exp(-\phi|\bm{s}_i-\bm{s}_j|)$, where $|\bm{s}_i-\bm{s}_j|$ is the distance
between $\bm{s}_i$ and $\bm{s}_j$, and $\phi$ is the range parameter for spatial
correlation. For the Gaussian process prior, the regression coefficients of
closer locations have strong correlation.

\subsection{Conditional Autoregressive Prior}
From \cite{banerjee2014hierarchical}, if the spatial domain $D$ is fixed and is
partitioned into a finite number of areal units $(A_1,\cdots,A_n)$ with
boundaries, data collected from such areal units, such as
cancer patients in counties of a state,
is known as areal data. For areal
data, the spatial association depends on neighborhood structures. Generally,
the
neighborhood structure for $n$ areal units comes from an $n \times n$
adjacency matrix $W$, where $w_{ij}=1$ if areal units $A_i$ and $A_j$ share
a common boundary, and 0 otherwise.
The conditional autoregressive model
\citep[CAR;][]{besag1974spatial} is one of the most popular tools to model
spatial correlations. It also has an advantage for being computationally
efficient for Gibbs sampling. For
Gaussian spatial random effects $\bm{\phi}=(\phi(A_1),\cdots,\phi(A_n))$, the
CAR model is defined as:
\begin{equation}
	\phi(A_i)\mid\phi(A_{-i}) \sim N(\rho\sum_{j\neq i} w_{ij}\phi(A_j),\tau^2),
	\label{car_full}
\end{equation}
where $\phi(A_{-i})=\{\phi(A_j),j\neq i\}$, $\tau^2$ is the conditional
variance, and $\rho\in \mathcal{R}$. Under the Brook's Lemma
\citep{brook1964distinction}, we can obtain the joint distribution of
$\bm{\phi}$ as:
\begin{equation}
p(\phi(A_1),\cdots,\phi(A_n))\propto
\exp\left\{-\frac{1}{2}\bm{\phi}'D^{-1}(I-\rho W)\bm{\phi}\right\},
	\label{eq:car_joint}
\end{equation}
where $D$ is diagonal matrix with $D_{ii}=\tau^2$.
Equation~\eqref{eq:car_joint}
suggests
that $\bm{\phi}$ follows a multivariate normal distribution with mean $\bm{0}$
and variance covariance matrix $\tau^2(I-\rho W)^{-1}$. This requires that $W$
be symmetric and $\tau^2(I-\rho W)^{-1}$ be positive definite. Let
$\lambda_1\leq \cdots \lambda_n$ be the eigenvalues of $W$. The eigenvalues sum
to 0 since $\text{trace}(W) = 0$ which tells us that $\lambda_1 < 0 <
\lambda_n$. The matrix $I-\rho W$ is positive definite if and only if $\rho \in
(1/\lambda_1, 1/\lambda_n)$. Based on the joint distribution of CAR model, we
have Conditional Autoregressive type prior for spatially varying coeffcients:
\begin{equation}\label{eq:carprior}
	\begin{split}
		\tilde{\bm{\beta}}_j\sim \mbox{MVN}(\bm{0},\tau^2_j(I-\rho W)^{-1}),\\
		\tau^2_j\sim \mbox{IG}(a,b),\\
		\rho \sim \text{Unif}(1/\lambda_1,1/\lambda_n),
	\end{split}
\end{equation}
where
$\tilde{\bm{\beta}}_j=(\tilde\beta_j(\bm{s}_1),\ldots,\tilde\beta_j(\bm{s}_n))
^\prime$, $j=1,\ldots,p$.

\subsection{Dirichlet Process Mixture Prior}
Within the Bayesian framework, Dirichlet process mixture model (DPMM) can link
response variable to covariates through cluster membership
\citep{molitor2010bayesian}. Formally, a probability measure $G$ following a
Dirichlet process \citep[DP;][]{ferguson1973bayesian} with a concentration
parameter $\alpha$ and a base distribution~$G_0$  is denoted by $G \sim
\text{DP}(\alpha, G_0)$ if
\begin{equation}
\label{eq:DP}
(G(A_1), \cdots, G(A_r)) \sim \text{Dirichlet}(\alpha G_0(A_1), \cdots, \alpha
G_0(A_r)),
\end{equation}
where $(A_1, \cdots, A_r)$ are finite measurable partitions of the space
$\Omega$.

Several different formulations can be used for determining the DP. In this
work, we use the stick-breaking construction proposed by
\cite{sethuraman1994constructive} for DP realization, which is given as
\[
\begin{split}
\theta_c \sim G_0, \hspace{0.1cm} G = \sum_{c=1}^{\infty} \pi_c
\delta_{\theta_c}(\cdot),\\
\pi_1 = V_1, \hspace{0.1cm} \pi_c = V_c \prod_{\ell<c}(1-V_\ell),\\
V_c \sim \text{Beta}(1, \alpha),
\end{split}
\]
where $\theta_c$ is the $c$th matrix consisting of the possible values for the
parameters of $G_0$, $\delta_{\theta_c}(\cdot)$ denotes a discrete probability
measure concentrated at $\theta_c$, and $\pi_c$ is the random probability
weight between 0 and 1.

Based on the DPMM, we can model the the spatially varying coefficients as
following the Dirichlet process gaussian mixture prior with $G_0$ being the
multivariate normal distribution :
\begin{equation}
	\begin{split}
\bm{\beta}_{z_i} \stackrel{\text{ind}} \sim \mbox{MVN}(\bm{\mu},
\bm{\Sigma}),\\
P(z_i = c\mid\bm{\pi}) = \pi_c,\\
\pi_1 = V_1, \hspace{0.1cm} \pi_c = V_c \prod_{\ell<c}(1-V_\ell),
\hspace{0.1cm} V_c \sim \text{Beta}(1, \alpha),\\
\tilde{\bm{\beta}}(\bm{s}_i)=\sum_{i=1}^\infty \pi_i\bm{\beta}_{z_i},
	\end{split}
	\label{eq:gaussian_dp}
\end{equation}
where $\bm{\mu}$ and $\bm{\Sigma}$ are hyper parameters for the multivariate
normal distribution.

\section{Bayesian Inference}\label{sec:binf}

In this section, we present the code used for computation using \pkg{nimble}
\citep{de2017programming}. Also, as we discussed in the introduction that
Gaussian, CAR, and DP can all be used as the prior for modeling the variation
in the coefficient vector and a choice needs to be made among these three,
the model selection criterion LPML is discussed.

\subsection{Bayesian Computation}

A nimble model consists of four major parts: model code, model constants, data,
and the initial values for MCMC. The model code is syntactically similar to the
\textsf{BUGS} language. As an illustration, we denote the number of locations
as $m$, number of observations per county as $n$, and dimension of the
covariate vector as $p = 3$. Take the model with CAR prior as an example.
First, the model is defined using the
\ttt{nimbleCode()} function:
\begin{knitrout}
\definecolor{shadecolor}{rgb}{0.969, 0.969, 0.969}\color{fgcolor}\begin{kframe}
\begin{alltt}
\hlstd{aft_car} \hlkwb{<-} \hlkwd{nimbleCode}\hlstd{(\{}
    \hlkwa{for} \hlstd{(i} \hlkwa{in} \hlnum{1}\hlopt{:}\hlstd{m) \{}
        \hlkwa{for} \hlstd{(j} \hlkwa{in} \hlnum{1}\hlopt{:}\hlstd{n) \{}
            \hlstd{logtime[i, j]} \hlopt{~} \hlkwd{dnorm}\hlstd{(mu[i, j], sigma[i])}
            \hlstd{censor[i, j]} \hlopt{~} \hlkwd{dinterval}\hlstd{(logtime[i, j], censortime[i, j])}
            \hlstd{mu[i, j]} \hlkwb{<-} \hlkwd{inprod}\hlstd{(beta[z[i],} \hlnum{1}\hlopt{:}\hlstd{p], X[}\hlnum{1}\hlopt{:}\hlstd{p, i, j])}

        \hlstd{\}}
        \hlstd{sigma[i]} \hlopt{~} \hlkwd{dinvgamma}\hlstd{(}\hlnum{1}\hlstd{,} \hlnum{1}\hlstd{)}
    \hlstd{\}}
    \hlstd{correlation[}\hlnum{1}\hlopt{:}\hlstd{m,} \hlnum{1}\hlopt{:}\hlstd{m]} \hlkwb{<-} \hlkwd{diag}\hlstd{(}\hlnum{1}\hlstd{, m)} \hlopt{-} \hlstd{b} \hlopt{*} \hlstd{W}
    \hlstd{b} \hlopt{~} \hlkwd{dunif}\hlstd{(low, high)}
    \hlkwa{for} \hlstd{(i} \hlkwa{in} \hlnum{1}\hlopt{:}\hlstd{p) \{}
        \hlstd{prec[i,} \hlnum{1}\hlopt{:}\hlstd{m,} \hlnum{1}\hlopt{:}\hlstd{m]} \hlkwb{<-} \hlstd{sigmabeta[i]} \hlopt{*} \hlstd{correlation[}\hlnum{1}\hlopt{:}\hlstd{m,} \hlnum{1}\hlopt{:}\hlstd{m]}
        \hlstd{beta[}\hlnum{1}\hlopt{:}\hlstd{m, i]} \hlopt{~} \hlkwd{dmnorm}\hlstd{(mu_beta[}\hlnum{1}\hlopt{:}\hlstd{m],} \hlkwc{prec} \hlstd{= prec[i,} \hlnum{1}\hlopt{:}\hlstd{m,} \hlnum{1}\hlopt{:}\hlstd{m])}
        \hlstd{sigmabeta[i]} \hlopt{~} \hlkwd{dgamma}\hlstd{(}\hlnum{1}\hlstd{,} \hlnum{1}\hlstd{)}
    \hlstd{\}}
\hlstd{\})}
\end{alltt}
\end{kframe}
\end{knitrout}

The 4th row of code defines that the logarithm of survival time for the $j$th
observation from the $i$th county follows normal distribution with mean
\ttt{mu[i, j]} and standard deviation \ttt{sigma[i]}, wich corresponds to
Equation~\eqref{eq:multimodel}. Next, due to the existence of censoring, the censoring
indicator \ttt{censor[i, j]} equals 1 if \ttt{logtime[i, j]} is right censored,
and 0 otherwise. The normal mean \ttt{mu[i,j]} is connected to the covariate
vector of the corresponding observation via \ttt{inprod(beta[z[i], 1:p], X[1:p, i, j])}.
For county $i$, its corresponding scale parameter is set to have an inverse
gamma distribution with shape 1 and scale 1. The correlation matrix is set to
be $I-b W$ as in \eqref{eq:car_joint} with $\rho$ replaced by \ttt{b}. The two
endpoints \ttt{low} and \ttt{high} correspond to $1/\lambda_1$ and
$1/\lambda_n$ in \eqref{eq:carprior}.

In the second part, we declare the data list for the model, which include the
logarithm of observed survival times, indicator for censoring, the independent
variables $X$, the adjacency matrix $W$, and the censor times.
\begin{knitrout}
\definecolor{shadecolor}{rgb}{0.965, 0.965, 0.965}\color{fgcolor}\begin{kframe}
\begin{alltt}
\hlstd{data} \hlkwb{<-} \hlkwd{list}\hlstd{(}
    \hlkwc{logtime} \hlstd{= logtime,}
    \hlkwc{censor} \hlstd{= censor,}
    \hlkwc{X} \hlstd{= X,}
    \hlkwc{adjacency} \hlstd{= W,}
    \hlkwc{censortime} \hlstd{= censortime}
\hlstd{)}
\end{alltt}
\end{kframe}
\end{knitrout}

Next we set the list of constant quantities in the model code. The quantities
\ttt{low} and \ttt{high} are obtained based on the adjacency structure of
Louisiana counties.
\begin{knitrout}
\definecolor{shadecolor}{rgb}{0.965, 0.965, 0.965}\color{fgcolor}\begin{kframe}
\begin{alltt}
\hlstd{constants} \hlkwb{<-} \hlkwd{list}\hlstd{(}
    \hlkwc{n} \hlstd{=} \hlnum{100}\hlstd{,}
    \hlkwc{m} \hlstd{=} \hlnum{64}\hlstd{,}
    \hlkwc{mu_beta} \hlstd{=} \hlkwd{rep}\hlstd{(}\hlnum{0}\hlstd{, m),}
    \hlkwc{p} \hlstd{=} \hlnum{3}\hlstd{,}
    \hlkwc{low} \hlstd{=} \hlopt{-}\hlnum{0.358}\hlstd{,}
    \hlkwc{high} \hlstd{=} \hlnum{0.175}
\hlstd{)}
\end{alltt}
\end{kframe}
\end{knitrout}

Finally, the initial values for parameters are assigned.
\begin{knitrout}
\definecolor{shadecolor}{rgb}{0.965, 0.965, 0.965}\color{fgcolor}\begin{kframe}
\begin{alltt}
\hlstd{inits} \hlkwb{<-} \hlkwd{list}\hlstd{(}
    \hlkwc{beta} \hlstd{=} \hlkwd{matrix}\hlstd{(}\hlnum{0}\hlstd{, m, p),}
    \hlkwc{b} \hlstd{=} \hlnum{0}\hlstd{,}
    \hlkwc{sigma} \hlstd{=} \hlkwd{rep}\hlstd{(}\hlnum{1}\hlstd{, m),}
    \hlkwc{sigmabeta} \hlstd{=} \hlkwd{rep}\hlstd{(}\hlnum{1}\hlstd{, p)}
\hlstd{)}
\end{alltt}
\end{kframe}
\end{knitrout}

With all four parts properly defined, \pkg{nimble} provides an one-line
implementation to invoke the MCMC engine, which includes setting the chain
length, burn-in, thinning, etc.:
\begin{knitrout}
\definecolor{shadecolor}{rgb}{0.965, 0.965, 0.965}\color{fgcolor}\begin{kframe}
\begin{alltt}
\hlstd{mcmc.out} \hlkwb{<-} \hlkwd{nimbleMCMC}\hlstd{(}
        \hlkwc{model} \hlstd{= aftModel,}
        \hlkwc{niter} \hlstd{=} \hlnum{20000}\hlstd{,}
        \hlkwc{nchains} \hlstd{=} \hlnum{1}\hlstd{,}
        \hlkwc{nburnin} \hlstd{=} \hlnum{5000}\hlstd{,}
        \hlkwc{thin} \hlstd{=} \hlnum{1}\hlstd{,}
        \hlkwc{monitors} \hlstd{=} \hlkwd{c}\hlstd{(}\hlstr{"b"}\hlstd{,} \hlstr{"sigma"}\hlstd{,} \hlstr{"beta"}\hlstd{,} \hlstr{"sigmabeta"}\hlstd{),}
        \hlkwc{summary} \hlstd{=} \hlnum{TRUE}
    \hlstd{)}
\end{alltt}
\end{kframe}
\end{knitrout}

The configuration above indicates that the MCMC results of parameters \ttt{b},
\ttt{sigma}, \ttt{beta}, and \ttt{sigmabeta}. One chain is ran for 20000
iterations with the first 5000 as burnin and without thinning. Therefore,
finally we obtain 15000 samples for each parameter.

\subsection{Bayesian Model Selection}
 A commonly used model comparison criterion, the Logarithm of the
Pseudo-Marginal Likelihood \citep[LPML;][]{ibrahim2013bayesian}, is applied to model
selection.
The LPML can be obtained through the Conditional Predictive Ordinate (CPO)
values. Let $Y_{(-i)} = \{Y_j: j = 1, \cdots, i-1, i+1, n\}$ denote the
observations with the $i$th subject response deleted. The CPO for the $i$th
subject is defined as:
\begin{equation}
	\label{eq:CPO}
\text{CPO}_i = \int f(Y_i\mid\tilde{\bm{\beta}}(s_i)\pi(w(s),
\tilde{\bm{\beta}(s)}, \sigma_y^2\mid Y_{(-i)}) \dd (w(s), \bm{\beta}(s),
\sigma_y^2),
\end{equation} 
where 
\[
    \pi(w(s), \bm{\beta}(s), \sigma_y^2\mid Y_{(-i)}) = \frac{\prod_{j \ne i}
f(y(s_j)\mid \bm{\beta}(s),w(s),\sigma_y^2)\pi(w(s), \bm{\beta}(s),
\sigma_y^2\mid Y_{(-i)})}{c(Y_{(-i)})},
\]
and $c(Y_{(-i)})$ is the normalizing
constant. Within the Bayesian framework, a Monte Carlo estimate of the CPO can
be obtained as:
\begin{equation}
	\label{eq:CPOest}
\widehat{\text{CPO}}_i^{-1} = \frac{1}{T} \sum_{t=1}^{T}
\frac{1}{f(Y_i\mid  \tilde{\bm{\beta}}_t(\bm{s}_i), \sigma(\bm{s}_i)},
\end{equation}
where $T$ is the total number of Monte Carlo iterations. An estimate of the
LPML can subsequently be calculated as:
\begin{equation}
	\label{eq:LPML}
	\widehat{\text{LPML}} = \sum_{i=1}^{N} \text{log}(\widehat{\text{CPO}}_i).
\end{equation}
Intuitively, a larger LPML indicates better fit to the data, and the corresponding
model is more preferred.

\section{Simulation}\label{sec:simu}

In this section, we present simulation studies for scenarios where there is no
spatial variation in the covariate effects, and where there is indeed
spatial variation in the covariate effects.
Information of the 64 Louisiana counties, including their centroids
and their adjacency structure, is used.
After obtaining the final parameter estimates and their 95\% highest posterior
density (HPD) intervals, we evaluate them using the following
four performance measures:
\[
\begin{split}
    \text{mean absolute bias (MAB)} & = \frac{1}{64}\sum_{\ell=1}^{64}
    \frac{1}{100} \sum_{r=1}^{100} \left|\hat{\beta}_{\ell, m,r} - 
    \beta_{\ell,m} \right|, \\
    \text{mean of mean squared error (MMSE)} & = \frac{1}{64}\sum_{\ell=1}^{64}
    \frac{1}{100}\sum_{r=1}^{100} \left(\hat{\beta}_{\ell,m,r} - \beta_{\ell,m}
    \right)^2, \\
    \text{mean standard deviation (MSD)} & = \frac{1}{64}\sum_{\ell=1}^{64}
    \sqrt{\frac{1}{99}\sum_{r=1}^{100} \left(\hat{\beta}_{\ell, m, r} -
    \bar{\hat{\beta}}_{\ell,m}\right)^2}, \\
    \text{mean coverage rate (MCR)} & = \frac{1}{64}\sum_{\ell=1}^{64}
    \frac{1}{100}\sum_{r=1}^{100} \text{1}\left(
    \beta_{\ell, m} \in 95\% \text{ HPD interval for replicate $r$}
    \right),
\end{split}
\]
where $\beta_{\ell,m}$ denotes the true value of the parameter for the $m$th
covariate in the $\ell$th county, $\bar{\hat{\beta}}_{\ell,m}$ is the average
of point estimates in the 100 replicates of simulation, 
$\beta_{\ell,m}$ denotes the true underlying parameter, and $1()$
denotes the
indicator function. In addition, the performance of DP prior in clustering the
counties is assessed with the Rand Index \citep[RI;][]{rand1971objective}, whose
value being close to 1 indicates good clustering result.
The code used is available at
GitHub~(\url{https://github.com/nealguanyu/Bayesian_AFT_SVC}).

\subsection{Simulation Without Spatially Varying Coefficients}\label{sec:null}

First we consider the scenario where there is no spatial variation in the
coefficients. Survival data are generated with $\beta = (0.6, 0.35, -0.5)^\top$.
Censoring times are generated independently from $\mbox{Exp}(1)$. Next, the
three models are fitted to the datasets.
For each of the 64 counties considered, three covariates are generated for 100
observations identically and independently from $N(0,1)$. Chain
lengths are set to 50,000 with the first 20,000 as burn-in. With the thinning
interval set to 10, a total of 3,000 posterior samples are obtained for each
replicate. The performance of Bayesian spatial AFT with the three aforementioned
priors is reported in Table~\ref{tab:simnull}. The average point estimates over
the 100 replicates and 64 counties are reported as well. For DP prior, the
maximum number of clusters is initially set to 20.

It turns out that under the no spatial variation scenario, models based on all
three priors give similar and rather accurate estimation results. The Gaussian
and CAR priors, as they allow each location to have their own set of parameter, give
relatively more volatile parameter estimates than the DP prior, as in all 100
replicates the 64 counties are identified to be in the same cluster, and the
parameter estimates are essentially coming from a model where all observations
are used. The MAB, MMSE, MSD of parameter estimates given by the model with DP
prior are much smaller than those given by the other two models. The MCR, as a
consequence, is lower than MCR for Gaussian and CAR prior, but is still close to
the 0.95 nominal value.

\begin{table}[tbp]
    \centering 
        \caption{Average point estimates as well as the four performance
        measures for estimation under the null scenario without spatially
        varying coefficients for the three priors.}\label{tab:simnull}
    \begin{tabular}{lcccccc}
        \toprule 
        Prior & Parameter & Point Estimate & MAB & MMSE & MSD & MCR \\
\midrule 
Gaussian & $\beta_1$ & 0.608 & 0.039 & 0.002 & 0.049 & 0.992 \\
   & $\beta_2$ & 0.374 & 0.044 & 0.003 & 0.050 & 0.978 \\
   & $\beta_3$ & -0.519 & 0.043 & 0.003 & 0.050 & 0.986 \\ [1ex]
CAR & $\beta_1$ & 0.565 & 0.085 & 0.011 & 0.099 & 0.944 \\
   & $\beta_2$ & 0.326 & 0.079 & 0.010 & 0.094 & 0.951 \\
   & $\beta_3$ & -0.473 & 0.082 & 0.011 & 0.097 & 0.944 \\ [1ex]
DP & $\beta_1$ & 0.598 & 0.011 & 0.0002 & 0.014 & 0.930 \\
   & $\beta_2$ & 0.357 & 0.011 & 0.0002 & 0.013 & 0.930 \\
   & $\beta_3$ & -0.509 & 0.013 & 0.0002 & 0.013 & 0.920 \\
   \bottomrule 
    \end{tabular}
\end{table}

\subsection{Simulation with Spatially Varying Coefficients}

We consider a smooth variation of coefficients over the counties of Louisiana.
The latitude and longitude of county centroids are obtained, and normalized to
have mean 0 and standard deviation 1. For county $\ell$, the coefficient vector
\begin{equation}\label{eq:alter1}
    \beta_\ell^\top = (0.6, 0.35, -0.5) + 0.1 \times (\mbox{longitude}_\ell +
    \mbox{latitude}_\ell), ~~\ell=1,\ldots, 64,
\end{equation}
where $\mbox{latitude}_\ell$ and $\mbox{longitude}_\ell$ denote the normalized
coordinates, respectively. Other data generation settings are consistent with
Section~\ref{sec:null}. Each of the three priors are fitted on the same 100
replicates of simulated data. Results are reported in Table~\ref{tab:simalter1}.
In another case, instead of having the variation pattern depend on longitude and
latitude of county centroids, we considered the case where there is a small
random term at each county, i.e, 
\begin{equation}\label{eq:repsilon}
    \beta_\ell^\top = (0.6, 0.35, -0.5) + \Delta \beta^\top,
\end{equation}
where $\Delta\beta \sim \mbox{MVN}(\bm{0}, 0.1 \bm{I}_3)$. Corresponding results
are presented in Table~\ref{tab:simalter2}.

\begin{table}[tbp]
    \centering 
        \caption{Performance measures for estimation under the alternative
        scenario \eqref{eq:alter1} for the three priors.}\label{tab:simalter1}
    \begin{tabular}{lcccccc}
        \toprule 
        Prior & Parameter & MAB & MMSE & MSD & MCR \\
\midrule 
Gaussian & $\beta_1$ & 0.042 & 0.003 & 0.051 & 0.988 \\
   & $\beta_2$ & 0.046 & 0.003 & 0.051 & 0.974 \\
   & $\beta_3$ & 0.045 & 0.003 & 0.052 & 0.983 \\ [1ex]
CAR & $\beta_1$ & 0.086 & 0.012 & 0.101 & 0.938 \\
   & $\beta_2$ & 0.080 & 0.010 & 0.095 & 0.944 \\
   & $\beta_3$ & 0.083 & 0.011 & 0.099 & 0.942 \\[1ex]
DP & $\beta_1$ & 0.074 & 0.010 & 0.074 & 0.397 \\
   & $\beta_2$ & 0.073 & 0.009 & 0.068 & 0.403 \\
   & $\beta_3$ & 0.073 & 0.009 & 0.068 & 0.403 \\
   \bottomrule 
    \end{tabular}
\end{table}

\begin{table}[tbp]
    \centering 
        \caption{Performance measures for estimation under the alternative
        scenario \eqref{eq:repsilon} for the three priors.}\label{tab:simalter2}
    \begin{tabular}{lcccccc}
        \toprule 
        Prior & Parameter & MAB & MMSE & MSD & MCR \\
\midrule 
Gaussian & $\beta_1$ & 0.065 & 0.007 & 0.060 & 0.932 \\
   & $\beta_2$ & 0.065 & 0.007 & 0.058 & 0.916 \\
   & $\beta_3$ & 0.066 & 0.007 & 0.060 & 0.919 \\ [1ex]
CAR & $\beta_1$ & 0.087 & 0.012 & 0.101 & 0.941 \\
   & $\beta_2$ & 0.081 & 0.010 & 0.095 & 0.952 \\
   & $\beta_3$ & 0.084 & 0.011 & 0.099 & 0.943 \\[1ex]
DP & $\beta_1$ & 0.086 & 0.011 & 0.037 & 0.214 \\
   & $\beta_2$ & 0.069 & 0.008 & 0.022 & 0.323 \\
   & $\beta_3$ & 0.073 & 0.008 & 0.025 & 0.267 \\
   \bottomrule 
    \end{tabular}
\end{table}

From Tables~\ref{tab:simalter1} and \ref{tab:simalter2}, it is not surprising
that the Gaussian and CAR priors still give rather credible estimation results.
The DP prior, however, despite having MAB, MMSE and MSD roughly on the same
scale, has much lower MCR, which is due to the fact that in order to detect
clustered covariate effects, we are limiting the maximum number of clusters to
20. With such specification yet a continuously varying parameter
surface~\eqref{eq:alter1} or a randomly varying parameter
surface~\eqref{eq:repsilon}, failure for the DP parameter estimates to cover
some of the true values is inevitable.

\begin{figure}[tbp]
    \centering 
    \includegraphics[width=0.7\textwidth]{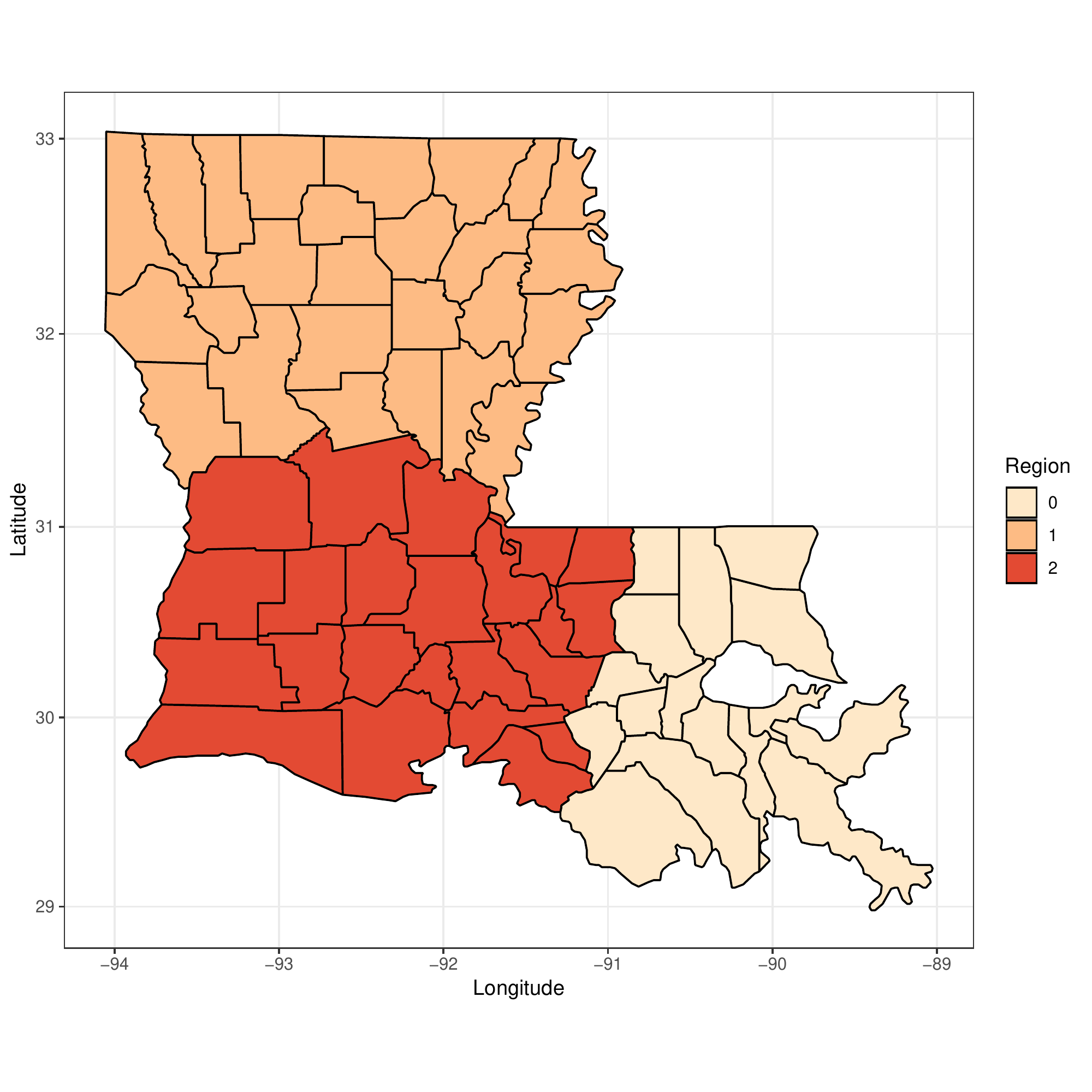}
    \caption{Illustration for the partition of Louisiana counties
    into three regions.}\label{fig:sim3}
\end{figure}

Similar to in \cite{ma2019geographically}, we also consider a setting where
counties in a region share the same covariate effects. The three-region
partition of Louisiana counties as illustrated in Figure~\ref{fig:sim3} is
considered. Under this setting, the RI for DP prior over the 100 replicates
averages to more than 0.972, indicating highly accurate clustering performance.
Compared to the Gaussian and CAR priors, the DP prior again yields parameter
estimates that are more stable, having much smaller MAB, MMSE, and MSD. As a
consequence of under-clustering in some replicates, i.e., less than three
clusters are identified, the MCR of DP prior is lower than the other two, but
still close to or higher than 0.85.

\begin{table}[tbp]
    \centering 
        \caption{Performance measures for estimation under the alternative
    regional scenario for the three priors.}\label{tab:simalter3}
    \begin{tabular}{lcccccc}
        \toprule 
        Prior & Parameter & MAB & MMSE & MSD & MCR \\
\midrule 
Gaussian & $\beta_1$ & 0.150 & 0.035 & 0.171 & 0.961 \\
   & $\beta_2$ & 0.187 & 0.051 & 0.152 & 0.893 \\
   & $\beta_3$ & 0.218 & 0.069 & 0.169 & 0.870 \\ [1ex]
CAR & $\beta_1$ & 0.187 & 0.055 & 0.222 & 0.954 \\
   & $\beta_2$ & 0.193 & 0.059 & 0.205 & 0.941 \\
   & $\beta_3$ & 0.216 & 0.073 & 0.202 & 0.906 \\[1ex]
DP & $\beta_1$ & 0.046 & 0.013 & 0.108 & 0.898 \\
   & $\beta_2$ & 0.042 & 0.011 & 0.103 & 0.898 \\
   & $\beta_3$ & 0.044 & 0.009 & 0.089 & 0.845 \\
   \bottomrule 
    \end{tabular}
\end{table}

\section{Survival Analysis of SEER Prostate Cancer Patients}\label{sec:apply}

\begin{figure}[tbp]
    \centering
    \includegraphics[width = \textwidth]{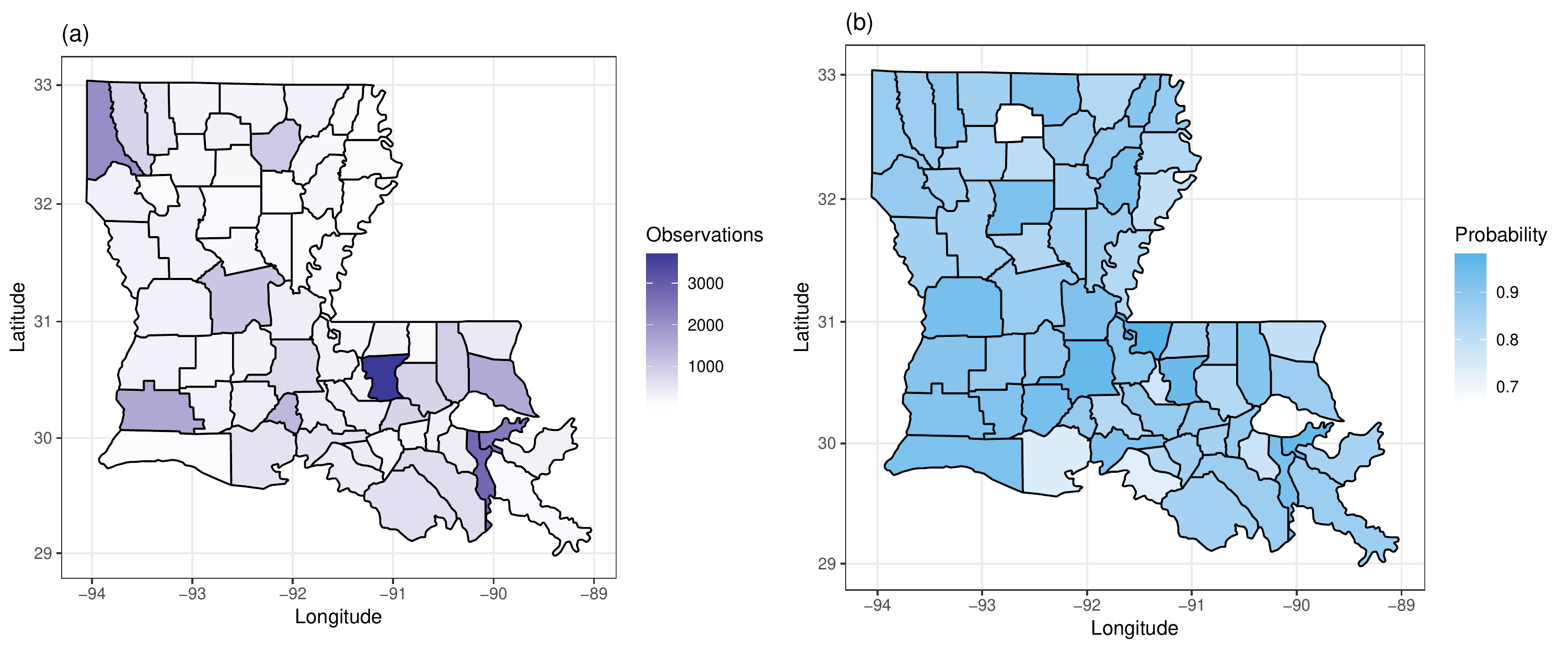}
    \caption{(a) Number of observations in counties of Louisiana; (b)
    Kaplan--Meier estimate of survival probability at 50 months after
diagnosis in each county.}\label{fig:demo}
\end{figure}

We use the dataset on prostate cancer patients from the SEER program as an
illustration for applicability of the proposed methods. There are 31,271
patients diagnosed with prostate cancer between 1973 to 2013, 2,057 of
which experienced events due to prostate cancer within the follow-up period,
resulting in a state-level censoring rate of 93.4\%. Three risk factors are
considered in our analysis: age at diagnosis (\textsf{Age}, centered and
scaled), marital status indicator (\textsf{Married}), and indicator for being
non-White (\textsf{Race}). Survival times are reported in integer months. For
those whose observed time is 0, a minor 0.0001 adjustment term is added to the
survival time to avoid negative infinity log times. The demographic
characteristics for this dataset is presented in Table~\ref{tab:demo}. In
Figure~\ref{fig:demo}, number of observations and per-county Kaplan--Meier (KM)
survival probability estimates at 50 months after diagnosis are plotted on the
map of Louisiana. Tensas county has the smallest number of observations (37),
and East Baton Rouge has the largest number of observations (3614). At 50 months
after diagnosis, the KM survival probability is highest for East Carroll
(0.972), and lowest for Allen (0.658).

\begin{table}[tbp]
  \centering 
  \caption{Demographics for the studied dataset. For continuous variables, the
  mean and standard deviation (SD) are reported. For binary variables, the
  frequency and percentage of each class are reported.} \label{tab:demo}
  \begin{tabular}{lc}
    \toprule 
     & Mean(SD) / Frequency (Percentage) \\ \midrule 
     Age & 64.78 (10.89)\\
     Survival Time & 63.77 (43.09) \\
     ~~Event & 40.02 (33.90) \\
     ~~Censor & 65.44 (43.17) \\
     Marital Status \\
     ~~Currently Married & 26 558 ($84.93\%$) \\
     ~~Other & 4 713 ($15.07\%$) \\
      Race \\
      ~~White & 21 674 ($69.31\%$)\\
      ~~Other & 9 597 ($30.69\%$)\\
      Cause-specific Death Indicator \\
      ~~Event & 2 057 ($6.58\%$) \\
      ~~Censor & 29 214 ($93.42\%$) \\
  \bottomrule 
  \end{tabular}
\end{table}

Estimation using each of the three priors discussed before is done on the
dataset. To determine which prior yields estimation that best suit the data, the
LPML values are calculated and reported in Table~\ref{tab:lpml}. As a larger
LPML value indicates a better fit, we choose to base our final conclusion on the
CAR prior model's results. In addition, to verify that covariates are indeed
varying, a vanilla model with no spatially varying effects is fitted, and it
turns out to have an LPML value of -595 240.9, indicating the existence of such
effects. In addition, it can be noticed that the DP prior based model has the
smallest LPML value, as it only identifies two clusters, and does not make the
model flexible enough.

\begin{figure}[tbp]
    \centering 
    \includegraphics[width = \textwidth]{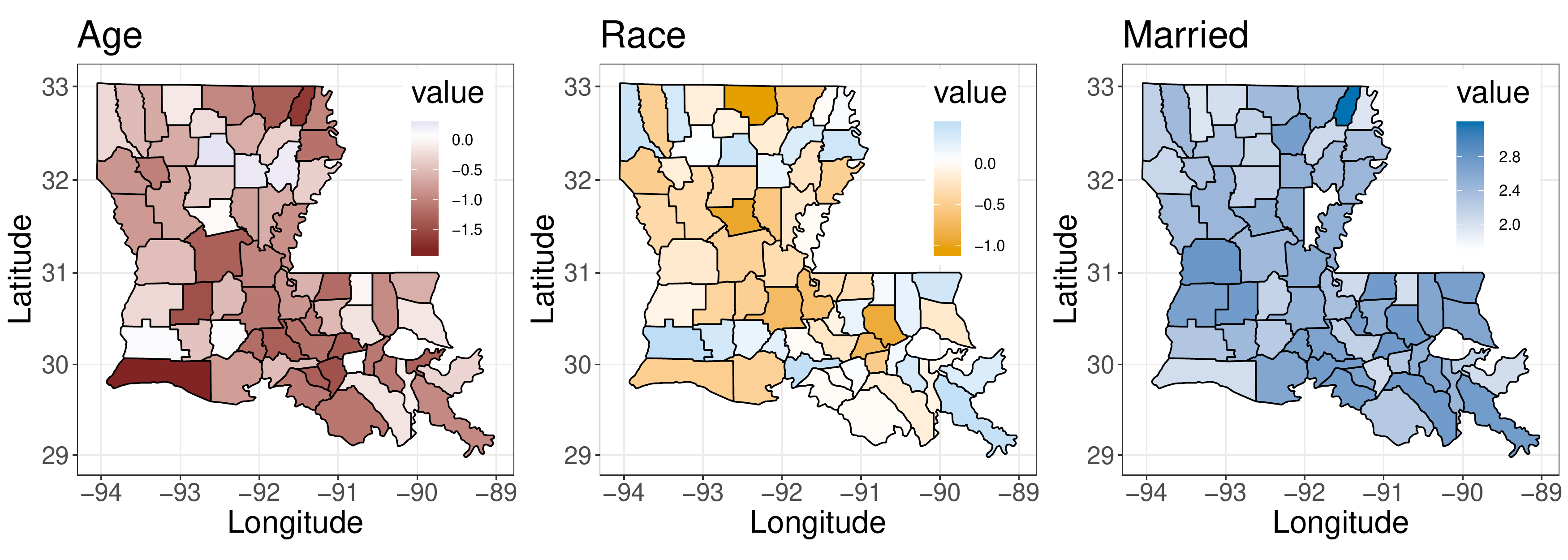}
    \caption{Visualization of final covariate effect estimates
    for Louisiana counties.}\label{fig:finalbetahat}
\end{figure}

Final estimation results are visualized on the county-level map of Louisiana in
Figure~\ref{fig:finalbetahat}. The overall pattern aligns with our intuition.
The parameter estimates for \textsf{Age} are negative in all counties,
indicating that older patients on average are more likely to experience an event
than younger patients. The parameter estimates for \textsf{Race} is also
negative in all counties, which suggests that there exist racial disparities in
the outcome of healthcare for prostate cancer in Louisiana. Finally, compared to
others, married patients are, on average, surviving for longer times. In
addition, the spatial variation in all three covariate effects is rather clear.

\begin{table}[tbp]
    \centering
    \caption{LPML values for different priors in modeling Louisiana data.}
    \label{tab:lpml}
    \begin{tabular}{cccc}
        \toprule 
        & Gaussian Process & CAR & DP \\ \midrule 
        LPML & -406 588.20 & -367 700.24 & -416 597.30 \\
        \bottomrule
        
    \end{tabular}
\end{table}

\section{Discussion}\label{sec:disc}

We proposed the usage of three different prior distributions in Bayesian
estimation of the spatially varying coefficients for the AFT model. The three
priors all have accurate performance under the null scenario where there is no
spatial variation, and are able to identify the varying patterns and produce
highly accurate parameter estimates under each of the three alternative
scenarios. In addition, when the spatial variation pattern is not smooth but
clustered, the DP prior is able to produce credible inference of cluster
belongings. The practical merit of the proposed method is illustrated using a
SEER prostate cancer data of patients in Louisiana.

A few issues are worth further investigating. In this work and many other
previous works in the spatially varying coefficients model context, oftentimes
all $p$ covariates are assumed to vary, which can lead to unnecessarily large
models if there are some coefficients are not varying. Identifying such
coefficients is an interesting topic. Extension of the proposed methods to the
proportional hazards model is also dedicated to future research.


%
%

\end{document}